\documentclass[twocolumn, prl, showpacs, floatfix]{revtex4}

\usepackage{amsmath}
\usepackage{graphicx}

\begin{document}

\title{Universality and non-universality in behavior of self-repairing random networks}
\author{A.~S.~Ioselevich, D.~S.~Lyubshin}
\affiliation{Landau Institute for Theoretical Physics, Russian
 Academy of Sciences, Kosygina str.2, 117940  Moscow, Russia,\\
Moscow Institute of Physics and Technology, Moscow 141700,
Russia.}
\date{\today}
\begin{abstract}
We numerically study one-parameter family of random single-cluster
systems. A finite-concentration topological phase transition from
the net-like to the tree-like phase (the latter is without a
backbone) is present in all models of the class. Correlation
radius index $\nu_B$ of the backbone in the net-like phase; graph
dimensions -- $d_{\min}$ of the tree-like phase, and $D_{\min}$ of
the backbone in the net-like phase appear to be universal within
the accuracy of our calculations, while the
 backbone fractal dimension $D_B$ is not universal: it depends on the parameter of a model.

\end{abstract}
\maketitle

Properties of disordered networks, such as porous materials, are
of great interest from both practical and theoretical point of
view (see \cite{book}). The standard percolation models (see
\cite{stauffer-aharony, bunde-havlin}) can not adequately describe
many of these systems, in particular the ``single-cluster'' ones,
which are obligatory connected (e.g., because of mechanical
stability requirements) and do not allow for detachment of finite
clusters. In our earlier paper \cite{srbp} we have introduced and
studied a simple model of self-repairing bond-percolation (SRBP)
in which all finite clusters arising in the process of gradual
destruction of randomly chosen bonds in the system are immediately
``repaired'' by means of regeneration of that very bond whose
destruction has lead to violation of connectivity. The system
therefore remains connected and constitutes a single large cluster
at all accessible concentrations. The SRBP-model turned out to be
analytically tractable by means of a partial mapping onto the
standard bond-percolation model; we were able to prove the
existence of a finite-density topological phase transition, at
which the backbone of the system vanishes and the system occurs in
a fragile tree-like state with zero mechanical rigidity and zero
electric conductivity. The backbone density $P_B$ and correlation
radius $\xi_B$ slightly above the transition follow the scaling
laws
\begin{eqnarray}
P_B(p)\propto (p-p_c)^{\beta_B},\qquad \xi_B(p)\propto
(p-p_c)^{-\nu_B} . \label{scaling1}
\end{eqnarray}
The critical concentration of bonds $p_c$ and relevant critical
exponents could be related to those of standard bond percolation
(though not being necessarily equal to them). In particular, in
two-dimensional lattices
\begin{eqnarray}
\beta_B^{\rm (SRBP)}\equiv \beta_B^{\rm
(perc)}=0.4757(10),\label{scaling1c}\\
 \nu_B^{\rm (SRBP)}\equiv
\nu^{\rm (perc)}= 4/3. \label{scaling1b}
\end{eqnarray}
The numerical value for $\beta_B^{\rm (perc)}$ in $2D$, given
in\eqref{scaling1c}, is taken from
 \cite{grassberger}, while the exact value of $\nu^{\rm (perc)}$ in $2D$
was first obtained in \cite{dennijs}. The fractal dimension of the
backbone is generally related to $\beta_B$ and $\nu_B$
\begin{eqnarray}
D_B=D-\beta_B/\nu_B. \label{scaling1e}
\end{eqnarray}
For the SRBP model it is also the same, as in percolation. In
$2D$:
\begin{eqnarray}
D_B^{\rm (SRBP)}\equiv D_B^{\rm
(perc)}=1.6432(8).\label{scaling1d}
\end{eqnarray}

 From the topological point of view the difference between the
tree-like phase and the usual net-like one is the abundant
presence of arbitrary large (up to the system size) cycles in the
graph of the net-like system and exponential decay of
concentration of large cycles on the graph of the tree-like
system.

One can easily envisage many other models of single-cluster random
networks. We were not able, however, to develop an analytical
approach to any physically reasonable model except SRBP. The
question about universality of the discovered phase transition
(and, moreover, the very existence of the transition in a
particular model) is very important and nontrivial. In contrast to
standard percolation models, the connectivity constraint (common
for all single-cluster models) is essentially nonlocal. This
increased complexity of the model makes the existence of relations
of single-cluster models to local field theories (similar to the
well-known relation between the standard percolation and the
Potts-model, see \cite{universality} and \cite{bunde-havlin})
highly improbable; we cannot therefore proclaim the universality
basing on standard renormalization-group arguments. Note, that the
invasion percolation (see \cite{invasion,bunde-havlin}), which
also involves a nonlocal constraint in its definition, does not
belong to the standard percolation universality class and is
characterized by its own critical indices. Thus any new
single-cluster model introduces a new puzzle of its own, and it is
very interesting and important to explore different patterns of
their behavior and try to establish some order in the
corresponding zoology. In this paper we start this program, and
study the non-universality of critical exponents.

Our first step was numerical exploration of the self-repairing
site-percolation (SRSP) on a square lattice, where randomly chosen
{\it sites} of the lattice (not {\it bonds}, as in SRBP) are
gradually removed and regenerated each time when the removal
causes disconnection of a finite cluster from the mainland.
Numerical simulation reveals a well-defined phase transition at
the site-concentration $x_c=0.608(1)$ with the critical index of
the backbone density $\beta_B^{\rm (SRSP)}=0.463(2)$, which is
distinct from the result \eqref{scaling1c} for the SRBP-model. The
latter difference is a definite manifestation of the
non-universality of the phase transition. On the other hand, our
simulation did not show any reliable difference between the SRSP
index of correlation radius: $\nu_B^{\rm (SRSP)}=1.337(8)$ and
$\nu^{\rm (SRBP)}=4/3$.  For the fractal dimension, using
\eqref{scaling1e}, we obtain
\begin{eqnarray}
D_B^{\rm (SRSP)}=1.653(2),\label{scaling1da}
\end{eqnarray}
also different from $D_B^{\rm (SRBP)}$.

 Another important fractal
characteristic of the critical backbone on the spatial scale
$R\ll\xi$ is the graph dimension $D_{\min}$, describing the
dependence of average "chemical distance" $\ell(R)$ (that is, the
length of the shortest path between two sites of a backbone,
separated by euclidean distance $R$)
\begin{eqnarray}
\ell_B(R)\propto R^{D_{\min}}. \label{scaling1a}
\end{eqnarray}
As we have shown in \cite{srbp},
\begin{eqnarray}
D_{\min}^{\rm (SRBP)}\equiv D_{\min}^{\rm (perc)}=1.13(2),
\label{scaling1af}
\end{eqnarray}
where the numerical value (for $2D$) was taken from
\cite{HerrmannStanley}. Note that the graph dimension of the
percolation backbone coincides with that of the entire infinite
cluster (see \cite{bunde-havlin}). Our simulations for the
SRSP-model give $D_{\min}^{\rm (SRSP)}=1.136(10)$, that, again, is
not different from the SRBP value \eqref{scaling1af} within our
accuracy.

\begin{figure}
\includegraphics[width=0.9\columnwidth]{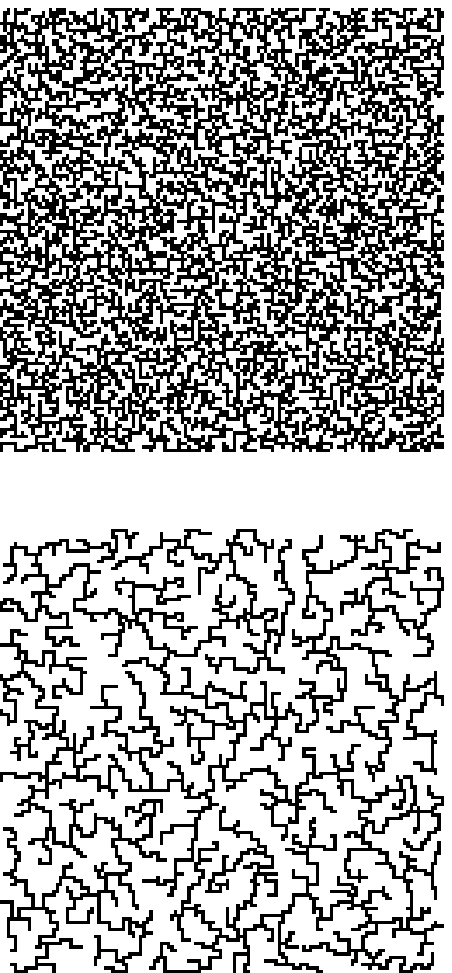}
\caption{SRSP-patterns in the tree-like phase: upper panel,
$x=0.5$, here few small cycles are still visible; lower panel,
$x=0.25$: concentration of cycles is already so low, that we do
not see any of them in this fragment of a quasi-tree.}
\label{srsp-0p0000-0p250-neg}
\end{figure}

 In contrast with the SRBP-model,
where the tree-like phase can only exist in the range $p_{\rm
tree}<p<p_c$, bounded from below by $p_{\rm tree}>0$ (see
\cite{srbp}), in SRSP-model the quasi-tree cluster can have
arbitrary low concentration ($0<x<x_c$). For $x\to 0$ the
concentration of cycles vanish very rapidly, so that the patterns
looks like a tree already slightly below $x_c$. In
Fig.\ref{srsp-0p0000-0p250-neg} reasonably large randomly chosen
fragments of a quasi-tree samples with $x=0.5$ and $x=0.25$ are
shown. It demonstrates that the cycles die out very fast with
lowering $x$: the  sample with $x=0.25$ is already practically a
tree.

The tree-like phase is fractal in the entire range $x<x_c$, so
that the chemical distance $\ell_T(R)$ in this phase obeys
\begin{eqnarray}
\ell_T(R)\propto R^{d_{\min}},\qquad  1<d_{\min}<2.
\label{scaling1a}
\end{eqnarray}
As was shown (both numerically and analytically) in \cite{srbp},
for SRBP-model $d_{\min}$ does not depend on $p$ throughout the
tree-like phase and exactly coincides with the similar index
$d_{\min}^{\rm (MST)}$ of the  Minimal Spanning Trees ensemble
(MST) on a lattice:
\begin{eqnarray}
d_{\min}^{\rm (SRBP)}\equiv d_{\min}^{\rm (MST)}=1.22(1).
\label{scaling1at}
\end{eqnarray}
The latter numerical value of $d_{\min}^{\rm (MST)}$ in $2D$ was
obtained in \cite{numerical dmin,manna} (see, also \cite{srbp}).

To determine $d_{\min}$  and $D_{\min}$ for SRSP model in this
paper we simulated systems of size $1024 \times 1024$ with open
boundary conditions. Statistics on average Euclidean displacement
$\langle R \rangle$ as a function of chemical distance $\ell$ was
gathered by performing breadth-first searches on points from the
central part of the system.  About $10^7$ starting points were
processed, and plain power fits $\langle R \rangle \sim
\ell^{1/d_{\min}}$ were applied in the stable region
$500<\ell<900$ not significantly affected by the boundary. Our
simulations give, again, $x$-independent
\begin{eqnarray}
d_{\min}^{\rm (SRSP)}=1.226(12), \label{scaling1ao}
\end{eqnarray}
throughout the tree-like range of $x$. This value, within our
accuracy,  does not differ from $d_{\min}^{\rm (MST)}$. To our
knowledge, however, there is no obvious relation between
quasitrees, appearing in the tree-like phase of the SRSP, and the
MST ensemble.

\begin{figure}
\includegraphics[angle=270,width=0.9\columnwidth]{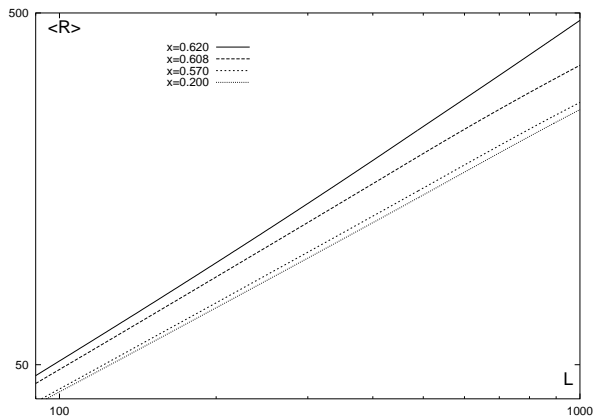}
\caption{Log-log plots for root mean square euclidean displacement
$R$ vs. chemical distance $\ell$
 for the SRSP-model on square lattice with four
different concentrations: $x=0.62$ (above the threshold);
$x=0.608$ (at the threshold); $x=0.57$ and $x=0.20$ (both below
the threshold). Within the accuracy of our calculations, the slope
of the first curve corresponds to $d_{\min}=1.0002$ (practically,
unity), the second curve -- to $D_{\min}=1.136(10)$. The  third
and the fourth curves both have the same slope, corresponding to
$d_{\min}=1.226(12)$ (practically, the same as $d_{\min}^{\rm
(MST)}$). }\label{srsp-dmin}
\end{figure}

\begin{figure}
\includegraphics[width=0.9\columnwidth]{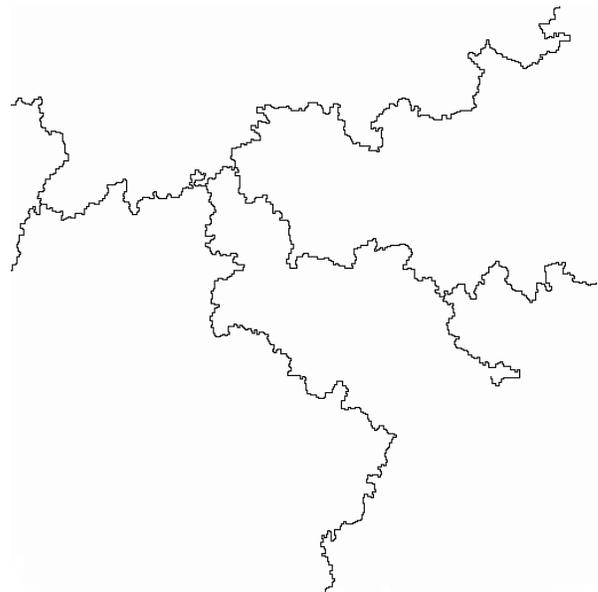}
\caption{SRSP-pattern at low density $x=0.0125$: not only a tree,
but apparently also a density fractal.} \label{low-density}
\end{figure}

At very low concentration $x\ll 1$ the tree-like phase of the
SR(S/B)P-models apparently acquire a new spatial scale
$\xi_C(x)\sim x^{-\nu_C}\gg 1$, -- the density correlation length,
that diverges as $x\to 0$. On the scales $1\ll L\ll\xi_C$ the
quasi-tree, besides being a
 "chemical fractal" (with nontrivial $d_{\min}>1$) becomes also
 a "density fractal" with nontrivial fractal dimension $D_C$ (see Fig.\ref{low-density}).
 Properties of this low-density
 phase and  corresponding critical indices will be studied in a
 separate publication.

\begin{figure}
\includegraphics[angle=270,width=0.9\columnwidth]{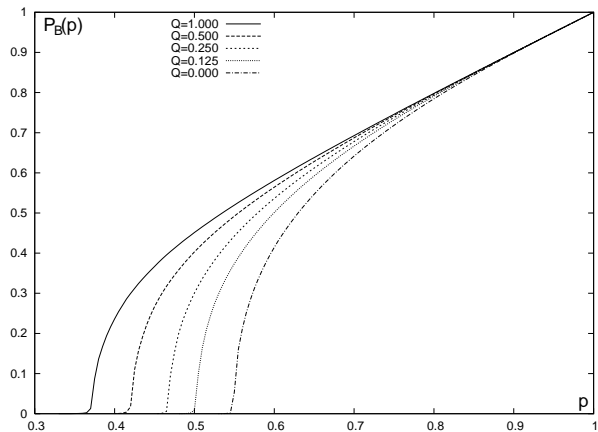}
\caption{The backbone density for SR(S/B)P hybrid models at
different values of mixing parameter $Q$. Concentration $p$ is
defined as a number of bonds present in the system; both $p$ and
$P_B$ are normalized by the total number of bonds in the full
lattice.
 }\label{srxp-backbone}
\end{figure}

 For further study of the non-universality we
have considered a one-parameter family of hybridized
SR(S/B)P-models, in which, at each step of the process, with
probability $1-Q$ a randomly chosen bond is removed (and restored,
if necessary) and, with probability $Q$, a randomly chosen site
together with all adjacent bonds is removed (and also restored, if
necessary). In Fig.\ref{srxp-backbone} the concentration
dependence of the backbone density $P_B(p)$ is shown for five
different values of the parameter $Q$. For three of them ($Q=0$,
$Q=0.5$, and $Q=1$) the critical exponents $\beta_B,\nu$ and
$\omega$ were accurately extracted from the data.
\begin{table}
\begin{tabular}{|c||c|c|c|c|}
\hline
$Q$ & $p_c$ & $\beta_B$ & $D_B$ & $\omega$ \\
\hline\hline
$0$  & $(3/4)(\sqrt{3}-1)$  & 0.4757(10)& 1.6432(8)& 1.85(1)\\
\hline
$0.5$ & 0.420476(1) & 0.468(2)& 1.649(2) & 1.785(5)\\
\hline
$1$  & 0.373116(1) & 0.463(2) & 1.653(2) & 1.788(3)\\
\hline
\end{tabular}
\caption{Nonuniversal characteristics of the family of SR(S/B)P
models.} \label{table1}
\end{table}
\begin{table}
\begin{tabular}{|c||c|c|c|}
\hline
$Q$  & $\nu_B$  & $d_{\min}$& $D_{\min}$\\
\hline\hline
$0$  & 4/3 & 1.22(1) & 1.13(1)\\
\hline
$0.5$ & 1.333(5)  & --- & ---\\
\hline
$1$   & 1.337(8) & 1.226(12)& 1.136(10)\\
\hline
\end{tabular}
\caption{Universal indices of the SR(S/B)P-family. }
\label{table2}
\end{table}

We have devised an efficient algorithm for simulations of models
with arbitrary $Q$ based on dynamic maintaining of connectivity on
the dual lattice that allowed site or bond deletion operations to
be performed in O(1) time. Lattices of size 128-256 turned out to
be the most useful, and $10^6$--$10^7$ samples were simulated for
each $Q$. This algorithm enabled us to accurately determine fine
differences in the critical exponents, appearing in the problem.

\begin{figure}
\includegraphics[angle=270,width=0.9\columnwidth]{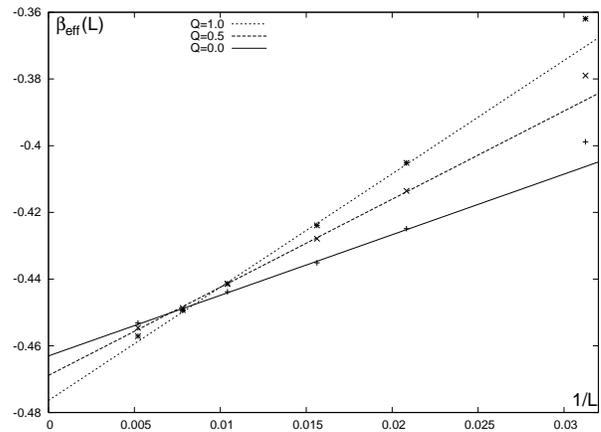}
\caption{Linear extrapolation of  $\beta_{\rm eff}(L)$ to $1/L\to
0$. The rescaling parameter was chosen $a=3/2$.
 }\label{srxp-beta}
\end{figure}

The procedure of finding the indices was as follows (c.f. paper
\cite{moukarzel}, where a similar procedure was proposed for study
of backbones for standard percolation): For a given $L$ an
ensemble of realizations on $L\times L$ square was generated with
an additional "boundary constraint", requiring that the sites and
bonds, belonging to  two opposite sides of the square (say, upper
and lower ones) could not be removed and constitute two "bars"
which by convention belong the the backbone. For a given
realization of the process the threshold $\tilde{p}_c$ was defined
as a concentration, at which the infinite cluster splits into two
parts -- one, connected to the upper bar, the other -- to the
lower bar. A distribution function of $\tilde{p}_c$ over the
ensemble was found. The average and the standard deviation of this
distribution are (see, e.g., \cite{stauffer-aharony}):
\begin{eqnarray}
\overline{p}_c(L)\equiv\overline{\tilde{p}_c}\approx
p_c+C_1L^{-\omega},\quad \Delta p_c(L)\approx C_2L^{-1/\nu_B},
\label{average1}
\end{eqnarray}
and the average value of the backbone density at the transition is
\begin{eqnarray}
\overline{P}_B(L)\equiv\overline{P_B(\tilde{p}_c)}\approx
C_3L^{D_B-D}, \label{average2}
\end{eqnarray}
where  $C_1$, $C_2$, $C_3$ -- critical amplitudes. The values of
$p_c$, $\nu_B$, $D_B$, $\beta_B$, and $\omega$, given in the
tables, were extracted from relations
(\ref{average1},\ref{average2}). The extraction procedure is
illustrated in Fig.\ref{srxp-beta}, where $\beta_{\rm
eff}(L)\equiv (\nu/\ln
a)\ln[\overline{P}_B(aL)/\overline{P}_B(L)]$, and $a>1$ is some
rescaling parameter (the result is not sensitive to the choice of
$a$). Note, that the exponent $\omega$ is {\it not universal}
already within the class of standard percolation (namely, it
depends on a particular lattice, as well as on the boundary
conditions), so that the dependence of $\omega$ on $Q$ may only be
viewed as an indirect evidence of non-universality of the SR(S/B)P
class.

In conclusion, we have studied a one parametric family  of
self-repairing site/bond percolation single-cluster models. In all
models of this family a topological phase transition between
net-like and tree-like phases was found, but  the backbone fractal
dimension $D_B(Q)$ turned out to be non-universal. It apparently
is a smooth monotonous function of the site/bond mixing parameter
$Q$, varying by $\sim 0.01$ in the range $0<Q<1$. Though the above
variation is quite small, it is well outside the error bars $\pm
0.002$ of our calculations. To reach this high accuracy special
efforts were made.

The non-universality of the critical indices was not unexpected,
since there are no known reasons for universality, similar to
those, existing for the family of standard percolation problems.
In this connotation, it was a great surprise that index $\nu_B$ of
the backbone correlation radius;  the graph dimension $D_{\min}$
of the backbone; and the graph dimension $d_{\min}$ of the
tree-like phase
 seemed to be $Q$-independent within our (rather high)
accuracy. Based on our numerical observations, we conjecture:
\begin{eqnarray*}
 \nu_B^{\rm (SR(S/B)P)}(Q)\equiv
\nu^{\rm (perc)}, \quad D_{\min}^{\rm (SR(S/B)P)}(Q,p)\equiv
D_{\min}^{\rm (perc)},\\
d_{\min}^{\rm (SR(S/B)P)}(Q,p)\equiv d_{\min}^{\rm (MST)}.
\end{eqnarray*}
So far, we did not find any rational explanation for this
intriguing phenomenon of "partial universality".

This work was supported by RFBR grant 06-02-16533.

\end{document}